\documentclass[10pt,letterpaper]{article}

\usepackage{cogsci}

\cogscifinalcopy 

\usepackage{pslatex}
\usepackage{apacite}
\usepackage{float} 

\usepackage{subcaption}
\usepackage{graphicx}
\graphicspath{{figures}}
\usepackage{makecell}
\usepackage{bm}
\usepackage{multirow}
\usepackage{footmisc}
\usepackage[table]{xcolor}
\usepackage{hhline}
\usepackage{wrapfig}
\usepackage{amsmath}

\newlength{\Oldarrayrulewidth}

\title{Bridging Declarative, Procedural, and  Conditional \\ Metacognitive Knowledge Gap Using  Deep Reinforcement Learning}

\author{{ \large \bf Mark Abdelshiheed, John Wesley Hostetter, Tiffany Barnes, and Min Chi} \\
  Department of Computer Science\\
  North Carolina State University\\
  Raleigh, NC 27695 \\
  \{{mnabdels,\, jwhostet,\, tmbarnes,\, mchi\}}@ncsu.edu}

\begin{document}

\maketitle

\begin{abstract}



In deductive domains, three metacognitive knowledge types in ascending order are declarative, procedural, and conditional learning. This work leverages Deep Reinforcement Learning (\textit{DRL}) in providing \textit{adaptive} metacognitive interventions to bridge the gap between the three knowledge types and prepare students for future learning across Intelligent Tutoring Systems (ITSs). Students received these interventions that taught \textit{how} and \textit{when} to use a backward-chaining (BC) strategy on a logic tutor that supports a default forward-chaining strategy. Six weeks later, we trained students on a probability tutor that only supports BC without interventions. Our results show that on both ITSs, DRL bridged the metacognitive knowledge gap between students and significantly improved their learning performance over their control peers. Furthermore, the DRL policy adapted to the metacognitive development on the logic tutor across declarative, procedural, and conditional students, causing their strategic decisions to be more autonomous.


\end{abstract}

\section{Introduction}

A demanding required feature of learning is being continuously prepared for future learning \shortcite{bransford1999transferRethinking}. Our incremental knowledge is the evidence that preparation for future learning exists yet is hard to predict and measure \shortcite{detterman1993transfer}. Considerable research has found that one factor that facilitates preparing students for future learning is their metacognitive knowledge \shortcite{kua2021MetacognitivePFL,cutrer2017MetacognitivePFL,hershkovitz2013MetacognitivePFL}.

Three types of metacognitive knowledge in deductive domains are \emph{declarative}, \emph{procedural}, and \emph{conditional} learning \shortcite{anderson2005ThreeTypes,schraw1998ThreeTypesMetacognition,bloom1956originalTaxonomy}. Although substantial research has shown that the three types could be acquired sequentially or simultaneously \shortcite{teng2020BridgingGap,yildirim2001BridgeGap,schraw1998ThreeTypesMetacognition}, it was also shown that each learner possesses a \emph{dominant} type of knowledge based on the educational context and learning environment \shortcite{kuhn2000metacognitiveDevelopment,brown1987ThreeTypesMetacognition}. Moreover, prior work has shown that students' metacognitive knowledge can evolve during learning \shortcite{rittle2001MetacognitiveDevelopment,baker1994metacognitiveDevelopment}. Thus, \emph{adaptive} interventions considering such development are needed \shortcite{azevedo2005adaptive,pintrich2002role,butler1998strategic}.

Reinforcement Learning (RL) \shortcite{sutton2018reinforcement} is one of the most effective approaches for \emph{adaptive} support and scaffolding across Intelligent Tutoring Systems (ITSs) \shortcite{krueger2017MetacognitiveRL,zhou2019hierarchical}. The deep learning extension of RL, known as deep RL (DRL), has been commonly utilized in pedagogical policy induction across ITSs \shortcite{abdelshiheed2023leveraging,alam2023exploring,hostetter2023leveraging,hostetter2023self,ju2021evaluating,sanz2019leveraging} due to its higher support of model sophistication. As far as we know,  no prior work has leveraged DRL in providing adaptive interventions to bridge the metacognitive knowledge gap and prepare students for future learning across ITSs.

This work builds on our prior work, where students were categorized into declarative, procedural, or conditional learners based on their \emph{dominant} metacognitive knowledge on an ITS. We found that only \textit{conditional} students were prepared for future learning, as they significantly outperformed their declarative and procedural peers across different deductive domains \shortcite{abdelshiheed2021preparing,abdelshiheed2020metacognition}. Inspired by such findings, this work empirically evaluates the DRL's effectiveness through a classroom study; we leverage DRL to provide \textit{adaptive} metacognitive interventions to bridge the knowledge gap for declarative and procedural students. DRL provided metacognitive interventions that teach students \textit{how} and \textit{when} to use a backward-chaining (BC) strategy on a logic tutor that supports a default forward-chaining (FC) strategy. After six weeks, we trained students on a probability tutor that only supports BC without receiving interventions. Our results showed that DRL indeed sealed the gap, prepared students for future learning, adapted to their metacognitive development, and raised their decision-making autonomy.


\section{Background \& Related Work}

\subsection{Declarative, Procedural and Conditional Knowledge}

Metacognition indicates cognition about cognition and the ability to control, conceive and regulate knowledge \shortcite{livingston2003metacognitionDefinition,roberts1993metacognitionDefinition,flavell1979metacognitionDefinition}.
Three types of metacognitive knowledge are declarative, procedural, and conditional \shortcite{schraw1998ThreeTypesMetacognition,schraw1995ThreeTypesMetacognition,brown1987ThreeTypesMetacognition,jacobs1987ThreeTypesMetacognition,bloom1956originalTaxonomy}. 

\textbf{Declarative} knowledge ---also described as surface or rote learning \shortcite{biggs1999teaching}--- is considered the simplest and lowest level of knowledge, as it involves memorization of facts and data offered in the default settings of a learning environment \shortcite{azevedo2013DeclarativeProcedural,de2012Declarative,krathwohl2002revisedTaxonomy,bloom1956originalTaxonomy}. \textbf{Procedural} knowledge is a higher form of knowledge that is discerned with the automated understanding of \textit{how} to use different problem-solving strategies and cognitive skills without conscious attention or reasoning about their rationale  \shortcite{azevedo2013DeclarativeProcedural,krathwohl2002revisedTaxonomy,dochy1992Procedural,georgeff1986procedural,bloom1956originalTaxonomy}. \textbf{Conditional} knowledge is the highest level of knowledge, as it requires understanding \textit{how}, \textit{when} and \textit{why} to use each strategy and cognitive skill \shortcite{kiesewetter2016ConditionalKnowledge,larkin2009ConditionalKnowledge,schraw1998ThreeTypesMetacognition}.

Considerable research has investigated the significance of acquiring and nurturing each knowledge type \shortcite{castro2021Conditional,boden2018DeclvsProc,fossati2009Procedural} and bridged the gap between them \shortcite{teng2020BridgingGap,mbato2019bridgeGap,yildirim2001BridgeGap}. \citeA{boden2018DeclvsProc} showed that students' self-efficacy was significantly correlated to declarative rather than procedural knowledge when solving physics problems. \citeA{fossati2009Procedural} found that students with high procedural knowledge significantly outperformed their peers on a tutoring system that teaches linked lists. In \citeA{castro2021Conditional}, students who had high conditional knowledge significantly surpassed their low peers in learning predictive parsing algorithms on a tutoring system.




\subsection{Metacognitive Development}

Metacognitive development is defined as the shifts in the learning approach and the metacognitive knowledge and skills used by a student \shortcite{anderson2005ThreeTypes,case2002metacognitiveDevelopment,rittle2001MetacognitiveDevelopment,kuhn2000metacognitiveDevelopment,baker1994metacognitiveDevelopment,anderson1982FrameworkDeclToProc}. We focus on the development across declarative, procedural, and conditional knowledge. \citeA{anderson1982FrameworkDeclToProc} proposed a framework for developing and acquiring metacognitive skills. He presented two major stages in skill acquisition: the declarative stage, where facts about the domain are interpreted, and the procedural stage, where domain knowledge is incorporated into procedures for performing the skill. \citeA{anderson1982FrameworkDeclToProc} stated that knowledge compilation occurs when the learner transitions from the declarative to the procedural stage. Later, in \citeA{anderson2005ThreeTypes}, he argued that procedural knowledge depends upon conditional knowledge, which in turn depends on declarative knowledge. Specifically, \citeA{anderson2005ThreeTypes} articulated that the learner's metacognitive knowledge develops from memorizing facts about strategies, then knowing the proper situations to use them, and finally, mastering each strategy.

\citeA{rittle2001MetacognitiveDevelopment} claimed that declarative and procedural knowledge develop in an iterative fashion through improved problem representations. They conducted two experiments on fifth- and sixth-graders learning decimal fractions and found that initial declarative knowledge predicted gains in procedural knowledge and vice versa. They showed that correct problem representation mediated the relation between declarative and procedural knowledge.

\subsection{Reinforcement Learning in ITSs}
Reinforcement Learning (RL) is a popular machine learning branch ideal in environments where actions result in numeric rewards without knowing a ground truth \shortcite{sutton2018reinforcement}. Due to its aim of maximizing the cumulative reward, RL has been widely used in educational domains due to the flexible implementation of reward functions \shortcite{gao2023hope,gao2022RL,ju2021evaluating,sanz2020exploring}. Deep RL (DRL) is a category of algorithms that combine RL algorithms with neural networks; for instance, Deep Q-Network (DQN) algorithm is the neural network extension of the Q-learning algorithm \shortcite{mnih2015DQN}. Substantial work has used RL and DRL in inducing pedagogical policies across ITSs \shortcite{ju2021evaluating,sanz2020exploring,zhou2019hierarchical}. \citeA{zhou2019hierarchical} utilized hierarchical RL to improve the learning gain on an ITS and showed that their policy significantly outperformed an expert and a random condition.

\citeA{ju2021evaluating} presented a DRL framework that identifies the critical decisions to induce a critical policy on an ITS. They evaluated their critical-DRL framework based on two success criteria: \textit{necessity} and \textit{sufficiency}. The former required offering help in \textit{all} critical states, and the latter required offering help \textit{only} in critical states. Their results showed that the framework fulfilled both criteria. \citeA{sanz2020exploring} conducted two consecutive classroom studies where DRL was applied to decide whether the student or tutor should solve the following problem. They found that the DRL policy with simple explanations significantly improved students' learning performance more than an expert policy.

Despite the wide use of RL and DRL on ITSs, the attempts to combine either with metacognitive knowledge have been minimal \shortcite{krueger2017MetacognitiveRL}.  \citeA{krueger2017MetacognitiveRL} used RL to teach the metacognitive skill of knowing how much to plan ahead (Deciding How to Decide). Their metacognitive RL framework builds on the semi-gradient SARSA algorithm developed to approximate Markov decision processes. They defined a meta Q-function that takes the meta state of the environment and the planning horizon action. They evaluated their framework on two planning tasks, where constrained reward functions were defined such that the rewards could be predicted many steps ahead to facilitate forming a plan. 

To sum up, despite much prior work on declarative, procedural, and conditional knowledge, it has yet to investigate the impact of closing the gap between them on preparation for future learning across ITSs. Our work utilizes DRL in providing adaptive metacognitive interventions to bridge the gap across ITSs. We investigate the impact of our interventions on students' metacognitive development and preparation for future learning. In brief, we induce and deploy a DRL policy of metacognitive interventions on a logic tutor and investigate its impact on a subsequent probability tutor.

\section{Logic and Probability Tutors}

\begin{figure}[ht!]
     \centering
     \begin{subfigure}[t]{0.194\textwidth}
         \centering
         \includegraphics[width=\textwidth,height=3.5cm]{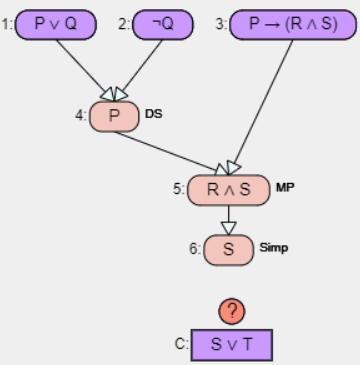}
         \caption{Forward Chaining (FC)}
         \label{fig:direct}
     \end{subfigure}
     \hfill
     \begin{subfigure}[t]{0.28\textwidth}
         \centering
         \includegraphics[width=\textwidth,height=3.5cm]{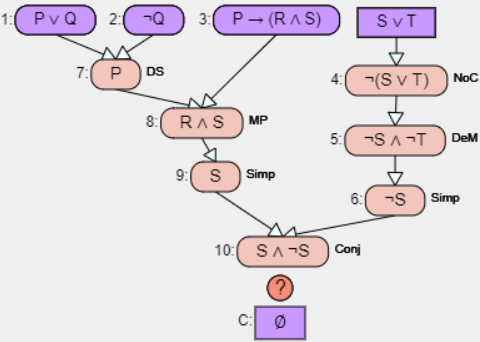}
         \caption{Backward Chaining (BC)}
         \label{fig:indirect}
     \end{subfigure}
\caption{Logic Tutor Problem-Solving Strategies}
\label{DT}
\end{figure}

\subsubsection{Logic Tutor}
It teaches propositional logic proofs by applying valid inference rules such as Modus Ponens through the standard sequence of pre-test, training and post-test. The three phases share the same interface, but training is the \emph{only} one where students can seek and get help. The pre-test has two problems, while the post-test is harder and has six problems; the first two are isomorphic to the pre-test problems. Training consists of five ordered levels with an \emph{incremental degree of difficulty}, and each level consists of four problems. Every problem has a score in the $[0,100]$ range based on the accuracy, time and solution length.

The \emph{pre-} and \emph{post-test} scores are calculated by averaging their pre- and post-test problem scores. A student can solve any problem throughout the tutor by either a \textit{FC} or a \textit{BC} strategy \shortcite{shabrina2023investigating,shabrina2023impact,shabrina2023learning}. Figure \ref{fig:direct} shows that for FC, one must derive the conclusion at the bottom from givens at the top, while Figure \ref{fig:indirect} shows that for \emph{BC}, students need to derive a contradiction from givens and the \emph{negation} of the conclusion. Problems are presented by \emph{default} in FC, but students can switch to BC by clicking a button in the tutor interface.

\begin{figure}[ht!]
\begin{center}
\includegraphics[width=0.34\textwidth]{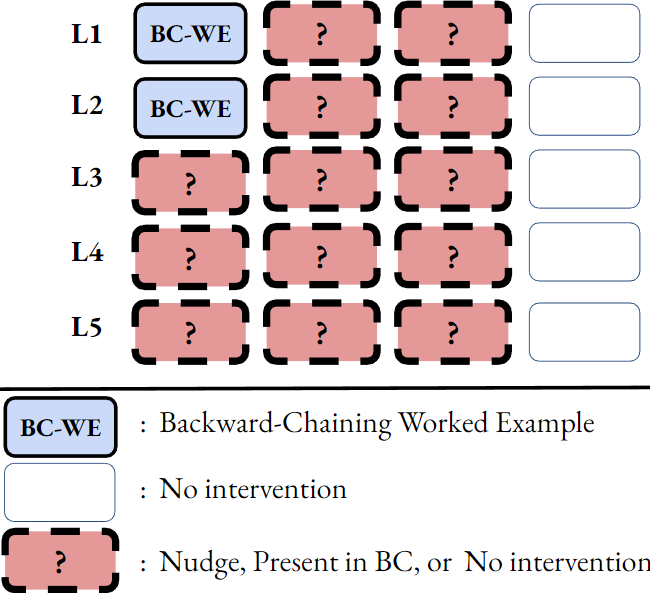}
\end{center}
\caption{Training on the Modified Logic Tutor (\textit{DRL})} 
\label{fig:modified_logic}
\end{figure}
\subsubsection{Probability Tutor}

It teaches how to solve probability problems using ten principles, such as the Complement Theorem. The tutor consists of a textbook, pre-test, training, and post-test. Like the logic tutor, training is the only section for students to receive and ask for hints, and the post-test is harder than the pre-test. The textbook introduces the domain principles, while training consists of $12$ problems, each of which can \emph{only} be solved by \textit{BC} as it requires deriving an answer by \emph{writing and solving equations} until the target is ultimately reduced to the givens. 

In pre- and post-test, students solve $14$ and $20$ open-ended problems, where each pre-test problem has an isomorphic post-test one. Answers are graded double-blind by experienced graders using a partial-credit rubric, where grades are based \emph{only} on accuracy in the $[0,100]$ range. The \emph{pre-} and \emph{post-test} scores are the average grades in their sections.

\section{Methods}

\noindent\textbf{Three Metacognitive Interventions }
As students can choose to switch problem-solving strategies \emph{only} on the logic tutor, our interventions are provided on the logic training. We previously found that \textit{\textbf{Conditional}} students frequently switched \emph{\textbf{early}} (\textbf{within the first $\mathbf{30}$ actions}) to \textit{BC} on the logic tutor, \textit{\textbf{Procedural}} students switched \emph{\textbf{late}} (\textbf{after the first $\mathbf{30}$ actions}), and their \textit{\textbf{Declarative}} peers made \textbf{no} switches and used the default strategy \shortcite{abdelshiheed2020metacognition}. It was also shown that providing metacognitive interventions that presented problems directly in BC or recommended switching to BC ---referred to as Nudges--- caused \textit{Declarative} and \textit{Procedural} students to catch up with their \textit{Conditional} peers \shortcite{abdelshiheed2022power,abdelshiheed2022mixing}.

\begin{figure}[bt!]
\begin{center}
\includegraphics[width=0.28\textwidth]{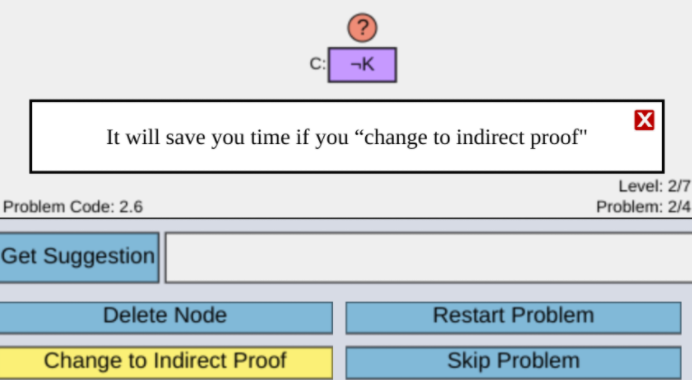}
\end{center}
\caption{Strategy Switch Nudge} 
\label{fig:prompt}
\end{figure}

This work leverages DRL to provide three metacognitive interventions regardless of the student's metacognitive group: \textit{\textbf{Nudge (Nud)}}, \textit{\textbf{Present in BC (Prs)}}, or \textit{\textbf{No Intervention (No)}}. We trained \textit{Experimental (DRL)} students on the modified tutor shown in Figure \ref{fig:modified_logic}. Two worked examples on BC were provided for teaching students \emph{how} to use BC, where the tutor showed a step-by-step solution.
Since our interventions included the no-intervention option, we intervened in as many problems as possible. Figure \ref{fig:prompt} shows an example of a nudge, which is prompted after a  number of seconds sampled from a probability distribution of prior students' switch behavior \shortcite{abdelshiheed2020metacognition}. We did not intervene in the last training problem of each level, as it is used to evaluate the improvement on that level.

\subsubsection{DRL Policy Induction}
We used data from four previous studies comprising $867$ students 
\shortcite{abdelshiheed2022power,abdelshiheed2022mixing,abdelshiheed2021preparing,abdelshiheed2020metacognition} and performed a $80-20$ train-test split. The dataset consisted of a record (\textbf{state}, \textbf{action}, \textbf{reward}) per student per logic training problem. The \textit{state} is the feature vector incorporating $152$ features that capture temporal, accuracy-based and hint-based behaviors. The \textit{action} is Nudge, Present in BC, or No Intervention. The \textit{reward} is the immediate problem score in the logic tutor. 


Our objective is to investigate whether DRL works with our metacognitive interventions \textbf{rather} than \textit{which} DRL algorithm is better with our interventions. We preferred DRL to RL due to its prevailing success in educational domains \shortcite{sanz2020exploring}. To select the algorithm, we had to avoid a relatively simple one such as Deep Q-Network (DQN), which overestimates action values \shortcite{mnih2015DQN} and may result in underfitting. Furthermore, we needed to avoid sophisticated DRL algorithms, such as autoencoders and actor-critic approaches, so that DRL does not overshadow the impact of our metacognitive interventions. In other words, a sophisticated DRL algorithm yielding an optimal policy would be acknowledged likely for its sophistication rather than for the metacognitive interventions it provided. Thus, we exploited Double-DQN (DDQN), which solves the overestimation issue in DQN by \textbf{decoupling} the action \textit{selection} from \textit{evaluation} in two different neural networks \shortcite{van2016DoubleDQN}. The resulting modified Bellman equation becomes:

\vskip -0.2in
\begin{equation}
\label{eq3}
     Q(s, a; \boldsymbol{\theta}) = r + \gamma \, \, Q(s', argmax_{{a'}} \,\, Q(s', a', \boldsymbol{\theta}); \boldsymbol{\theta^-})
\end{equation}

\noindent where $r$ is the reward; $\gamma$ is the discount factor; $s$ and $s'$ refer to the current and next states; $a$ and $a'$ denote the current and next actions. DDQN uses the \textbf{main} $(\boldsymbol{\theta})$ neural network to \textit{select} the action with the highest Q-value for the next state and then \textit{evaluates} its Q-value using the \textbf{target} $(\boldsymbol{\theta^-})$ neural network. After hyperparameter tuning, we picked the model with the lowest mean squared error loss. The deployed policy had two hidden layers with $16$ neurons each, $1e$-$3$ learning late, $9e$-$1$ discount factor, $32$ batch size, a synchronization frequency of $4$ steps between both neural networks ($\boldsymbol{\theta}$ and $\boldsymbol{\theta^-}$), and was trained until convergence ($\approx 2000$ epochs).

\section{Experiment Setup}


The experiment took place in an undergraduate Computer Science class in the Fall of 2022 at North Carolina State University. The participants were assigned each tutor as a class assignment and told that completion is required for full credit. We randomly split students into \textit{Experimental (DRL)} and \textit{Control (Ctrl)} conditions, where students were first assigned the logic tutor and then the probability tutor six weeks later. On both tutors, students received the problems in the same order and followed the standard phases described in the Logic and Probability Tutors section. The \textit{only} difference between the two conditions is that \textit{DRL} students received our \textit{adaptive} metacognitive interventions in the \textbf{logic training} provided by the DRL policy (Fig. \ref{fig:modified_logic}), while their \textit{Ctrl} peers received no interventions. On probability, all students received no interventions. 

The main challenge in this work was to ensure an even distribution of the metacognitive groups \{\textit{Declarative (Decl)}, \textit{Procedural (Proc)}, \textit{Conditional (CDL)}\} across the two conditions \{\textit{DRL}, \textit{Ctrl}\}. We aimed to investigate whether DRL would help students with different incoming metacognitive knowledge. Therefore, as per our prior work, we utilized the random forest classifier (RFC) that, based on pre-test performance, predicts the incoming metacognitive group before training on logic and was previously shown to be $96\%$ accurate \shortcite{abdelshiheed2021preparing}.

A total of $112$ students finished both tutors. We found that our DRL policy provided no interventions for \textit{CDL}$_{DRL}$ students $94\%$ of the time. Therefore, we combined \textit{CDL}$_{DRL}$ and \textit{CDL}$_{Ctrl}$ into \textit{CDL}. After randomly splitting students into conditions and utilizing the RFC for even distribution, we had $22$ \textit{Decl}$_{DRL}$, $24$ \textit{Proc}$_{DRL}$, $22$ \textit{Decl}$_{Ctrl}$, $22$  \textit{Proc}$_{Ctrl}$ and $22$ \textit{CDL} students. The RFC was $98\%$ accurate in classifying students who received no interventions ---\textit{Decl}$_{Ctrl}$, \textit{Proc}$_{Ctrl}$ and \textit{CDL}.

\section{Results}

\subsection{Learning Performance}

\begingroup
\renewcommand{\arraystretch}{1.5}
\begin{table}[ht!]
\scriptsize
\begin{center} 
\caption{Comparing Groups across Tutors} 
\label{exp-control-conditional} 
\begin{tabular}{ccc|c} 
\Xhline{4\arrayrulewidth}

\makecell{}   & \makecell[t]{\textit{Experimental} \\ $($DRL$:N=46)$} & \makecell[t]{\textit{Control} \\ $($Ctrl$:N=44)$} & \makecell[t]{\textit{Conditional} \\ $($CDL$:N=22)$}  \\
\hline
\multicolumn{4}{c}{Logic Tutor}\\
\hline
\textit{Pre} &  $55.9\,(21)$ & $55.8\,(21)$  & \cellcolor{gray!30}$58.2\,(19)$ \\
\textit{Iso. Post}  & $\mathbf{92.1\,(5)^{*}}$  & $73.4\,(17)$ & \cellcolor{gray!30} $83.4\,(12)^{*}$ \\
\textit{Iso. NLG} & $\mathbf{0.47\,(.1)}^{*}$ & $0.16\,(.28)$ & \cellcolor{gray!30} $0.35\,(.11)^{*}$ \\
\textit{Post} & $\mathbf{87.7\,(6)}^{*}$ & $69.8\,(15)$ & \cellcolor{gray!30} $80.2\,(11)^{*}$ \\
\textit{NLG}  & $\mathbf{0.45\,(.11)}^{*}$ & $0.13\,(.33)$ & \cellcolor{gray!30} $0.31\,(.15)^{*}$ \\

\hline
\multicolumn{4}{c}{Probability Tutor}\\
\hline
\textit{Pre} &  $75.7\,(15)$  &  $76\,(14)$ & \cellcolor{gray!30} $78.6\,(14)$ \\
\textit{Iso. Post}  & $\mathbf{95.3\,(4)^{*}}$ & $72.7\,(13)$ & \cellcolor{gray!30} $89.1\,(7)^{*}$\\
\textit{Iso. NLG} & $\mathbf{0.4\,(.12)}^{*}$  & -$0.04\,(.18)$ & \cellcolor{gray!30} $0.24\,(.15)^{*}$ \\
\textit{Post} & $\mathbf{94.9\,(4)^{*}}$  & $70.2\,(15)$ & \cellcolor{gray!30} $87.7\,(8)^{*}$\\
\textit{NLG}  &  $\mathbf{0.37\,(.14)}^{*}$  & -$0.08\,(.21)$ & \cellcolor{gray!30} $0.22\,(.19)^{*}$ \\

\Xhline{4\arrayrulewidth}
\end{tabular} 
\end{center} 
 {\centering In a row, bold is for the highest value, and asterisk means significance over no asterisks.\par}
\end{table}
\endgroup

\subsubsection{Experimental vs. Control vs. Conditional} Table \ref{exp-control-conditional} shows the groups' performance on both tutors. We display the mean and standard deviation of pre- and post-test scores, isomorphic scores, and the normalized learning gain (\textit{NLG}) defined as $(NLG = \frac{Post - Pre}{\sqrt{100 - Pre}})$, where $100$ is the maximum score. We refer to pre- and post-test scores as \textit{Pre} and \textit{Post}, respectively, while the groups are abbreviated as \textit{DRL}\footnote{Italicized version refers to group; otherwise refers to policy.}, \textit{Ctrl} and \textit{CDL}. We performed a Shapiro-Wilk normality test for each metric for each group and found no evidence of non-normality ($\mathit{p} >.05$). A one-way ANOVA using group as factor found no significant difference on \textit{Pre}: $\mathit{F}(2,109) = 0.09,\, \mathit{p} = .91$ for logic and $\mathit{F}(2,109) = 0.21,\, \mathit{p} = .81$ for probability. To measure the improvement on isomorphic problems, repeated measures ANOVA tests were conducted (one for each group on each tutor) using \{\textit{Pre}, \textit{Iso. Post}\} as factor. On both tutors, we found that \textit{DRL} and \textit{CDL} learned significantly with $\mathit{p} <.0001$, while \textit{Ctrl} did not ($\mathit{p} >.05$).

A one-way ANCOVA\footnote{General effect size $\eta^2$ was reported for conservative results.} using \textit{Pre} as covariate and group as factor found a significant effect on \textit{Post} on both tutors: $\mathit{F}(2,108) = 38.4,\, \mathit{p} < .0001, \, \mathit{\eta}^2 = 0.71$ for logic and $\mathit{F}(2,108) = 49.6,\, \mathit{p} < .0001, \, \mathit{\eta}^2 = 0.79$ for probability. Subsequent Bonferroni-corrected $(\alpha=.05/3)$ analyses revealed that \textit{DRL} significantly outperformed both groups, while \textit{CDL} significantly surpassed \textit{Ctrl}; for instance, \textit{DRL} had significantly higher \textit{Post} than \textit{CDL}: $\mathit{t}(66) = 3.6,\, \mathit{p} < .001$ and $\mathit{t}(66) = 3.8,\, \mathit{p} < .001$ for logic and probability, respectively. Similar results were found using ANOVA on \textit{NLG}. In brief, these findings on both tutors confirm that \textit{DRL} $>$ \textit{CDL} $>$ \textit{Ctrl}.

\subsubsection{Groups within Experimental and Control} 
\begin{figure}[ht!]
\begin{center}
\includegraphics[width=0.47\textwidth]{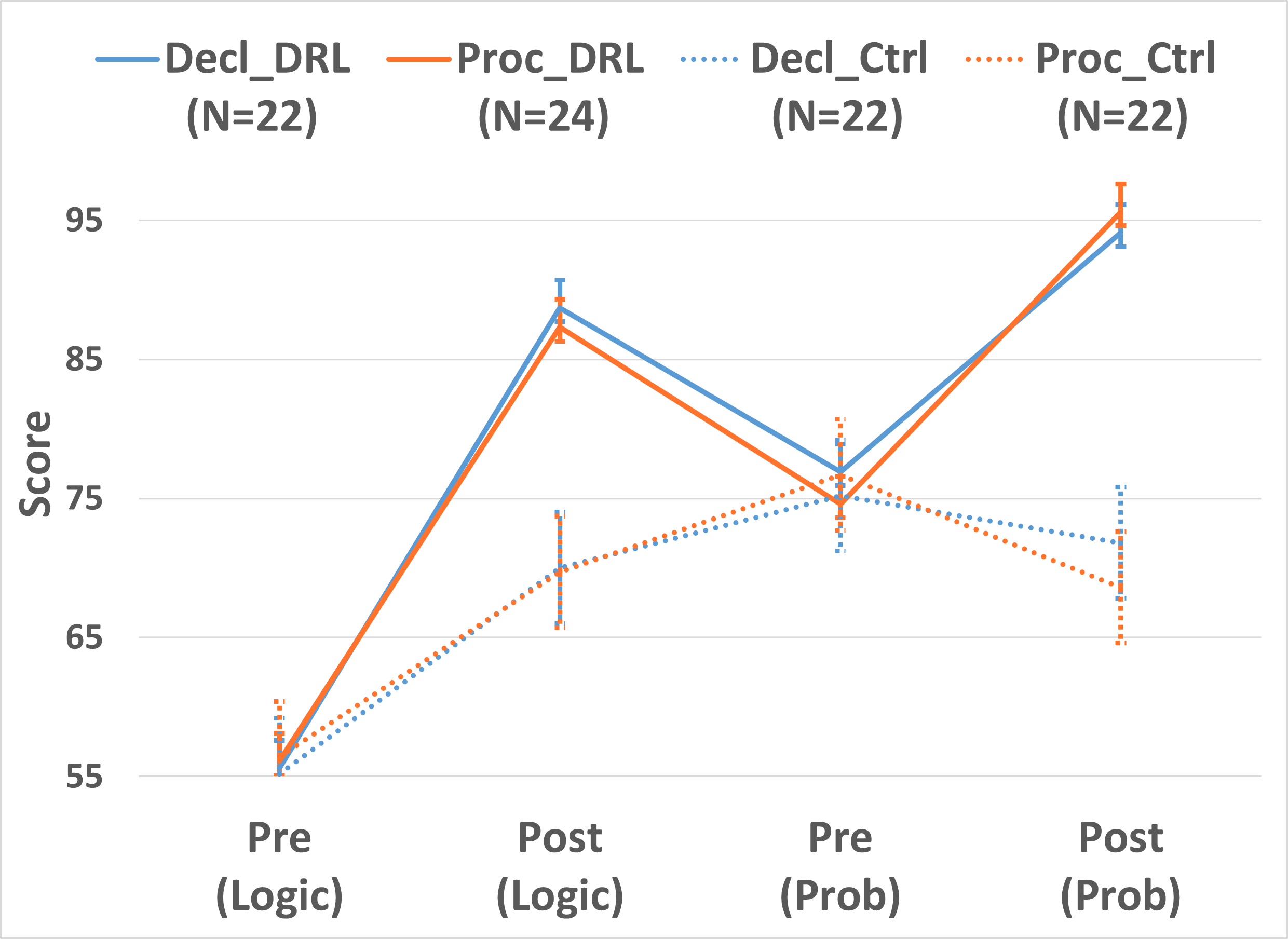}
\end{center}
\caption{Conditions \textit{Pre} and \textit{Post} Performance across Tutors} 
\label{fig:prepostPerformance}
\end{figure}

We compared the performance of the metacognitive groups \{\textit{Decl}, \textit{Proc}\} across the two conditions \{\textit{DRL}, \textit{Ctrl}\} to assess the within- and between-condition impact of the DRL policy. Our results showed the same statistically significant pattern on both tutors for \textit{Post}, \textit{NLG} and their isomorphic versions: \textit{Decl}$_{DRL}$, \textit{Proc}$_{DRL} >$ \textit{Decl}$_{Ctrl}$, \textit{Proc}$_{Ctrl}$. We display \textit{Pre} and \textit{Post} scores in Figure \ref{fig:prepostPerformance} and report the statistical results for \textit{NLG} in the following paragraph.

\noindent A two-way ANOVA using metacognitive group and condition as factors found a significant interaction effect on \textit{NLG} on both tutors: $\mathit{F}(1,86) = 36.3,\, \mathit{p} < .0001, \, \mathit{\eta}^2 = 0.68$ for logic and $\mathit{F}(1,86) = 45.8,\, \mathit{p} < .0001, \, \mathit{\eta}^2 = 0.76$ for probability. Follow-up Bonferroni-adjusted $(\alpha=.05/6)$ analyses showed that \textit{DRL} significantly outperformed \textit{Ctrl}. Specifically, on logic, \textit{Decl}$_{DRL}$ $[0.45(.12)]$ had significantly higher \textit{NLG} than \textit{Decl}$_{Ctrl}$ $[0.16(.33)]$: $\mathit{t}(42) = 6.2,\, \mathit{p} < .0001$, and \textit{Proc}$_{DRL}$ $[0.44(.14)]$ significantly outperformed \textit{Proc}$_{Ctrl}$ $[0.1(.31)]$: $\mathit{t}(44) = 7.3,\, \mathit{p} < .0001$. On probability, we found the same patterns, as \textit{Decl}$_{DRL}$ $[0.34(.13)]$ and \textit{Proc}$_{DRL}$ $[0.39(.16)]$ surpassed \textit{Decl}$_{Ctrl}$ $[-0.07(.32)]$ and \textit{Proc}$_{Ctrl}$ $[-0.1(.25)]$, respectively: $\mathit{t}(42) = 6.5,\, \mathit{p} < .0001$ and  $\mathit{t}(44) = 7.9,\, \mathit{p} < .0001$.

\subsection{Policy Decisions and Students' Strategic Behavior}

As the learning performance results found no significant difference within the \textit{DRL} group on both tutors, we further investigated the distribution of the DRL policy decisions; \textit{Decl}$_{DRL}$ received $94 (33\%)$ Nudges, $65 (23\%)$ presented in BC and $127 (44\%)$ No Intervention, while \textit{Proc}$_{DRL}$ received $82 (26\%)$ Nudges, $74 (24\%)$ presented in BC and $156 (50\%)$ No Intervention. A chi-square test showed no significant difference in the decisions' distribution between the \textit{DRL} groups:
$\chi^2 (2,\, N=598) = 3.2, \, \mathit{p}=.2$. Thus, we combined their policy decisions and analyzed the decisions' distribution per logic training level, as shown in Table \ref{policyDecisions}.

A chi-square test showed a significant relationship between the policy decision type and the training level: $\chi^2 (8,\, N=598) = 81.2, \, \mathit{p}<.0001$. Post-hoc
pairwise chi-square tests with Bonferroni adjustment $(\alpha=.05/10)$ showed that the last two levels had significantly more No-Intervention decisions. For instance, the fourth level had more No-Intervention decisions than the third level: $\chi^2 (2,\, N=276) = 33.4, \, \mathit{p}<.0001$.


\begingroup
\renewcommand{\arraystretch}{1.1}
\begin{table}[ht!]
\begin{center} 
\caption{Distribution of DRL Policy Decisions across Levels} 
\label{policyDecisions} 
\begin{tabular}{c|c|c|c|c|c} 
\Xhline{4\arrayrulewidth}

 & $L1$ ($\%$) & $L2$ ($\%$) & $L3$ ($\%$) & $L4$ ($\%$) & $L5$ ($\%$)\\
\hline
Nud & $\mathbf{40}$ & $\mathbf{42}$ & $\mathbf{37}$ & $17$ & $18$ \\

Prs & $28$ & $31$ & $31$ & $16$ & $15$ \\

No & $32$ & $27$ & $32$ & $\mathbf{67}$ & $\mathbf{67}$\\

\Xhline{4\arrayrulewidth}
\end{tabular} 
\end{center} 
 {\centering In a column (training level), bold is for the highest value.\par}
\end{table}
\endgroup

\begin{figure}[ht!]
\begin{center}
\includegraphics[width=0.48\textwidth]{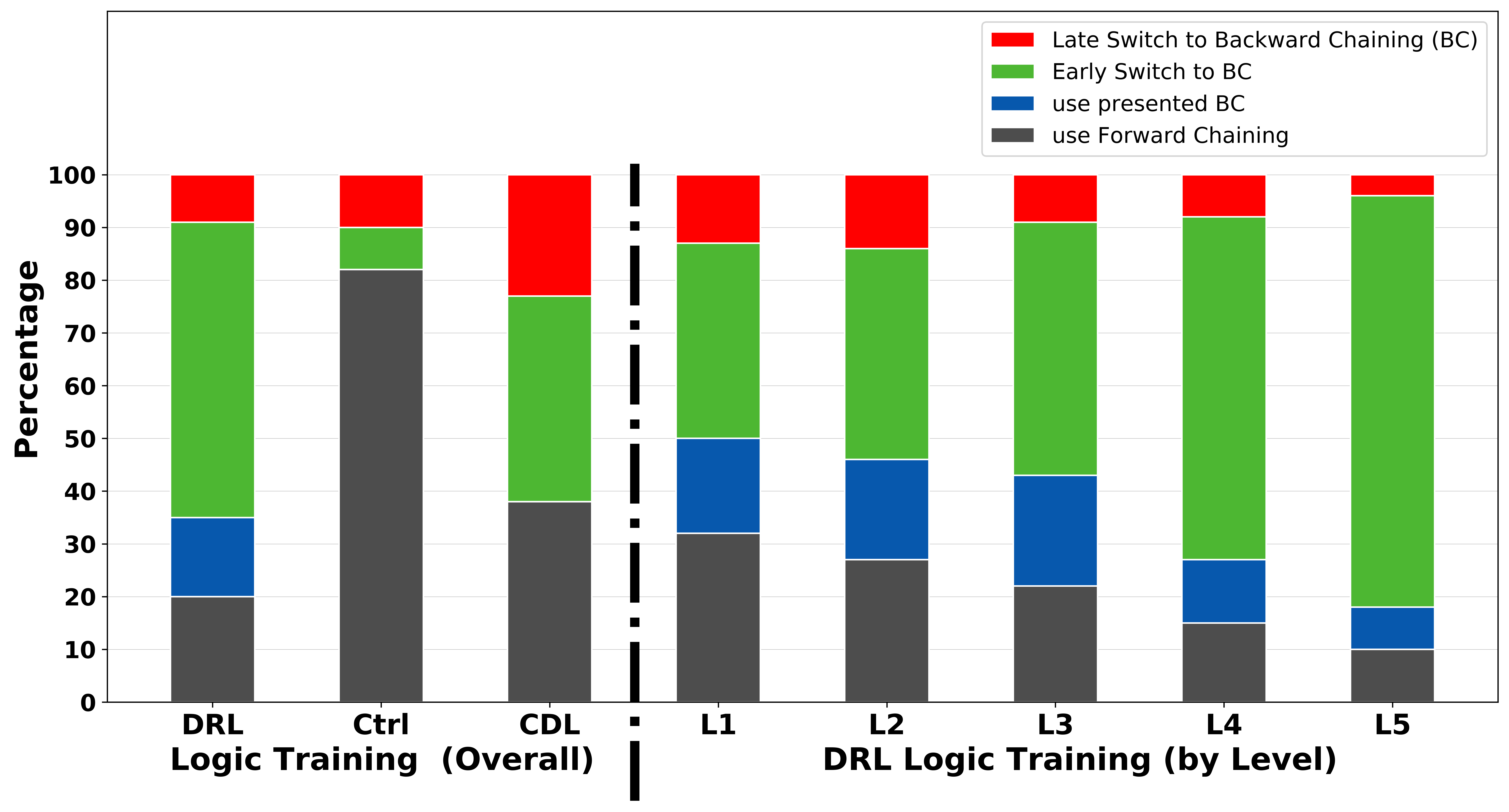}
\end{center}
\caption{Students' Strategic Decisions} 
\label{fig:switch}
\end{figure}

\noindent The students' strategic decisions on logic training are shown in Figure \ref{fig:switch} to investigate the impact of the DRL policy on their choices. The first three columns reflect the choices \textit{DRL}, \textit{Ctrl} and \textit{CDL} students made during the entire training, where \textit{Decl}$_{DRL}$ and \textit{Proc}$_{DRL}$ decisions had similar distributions and thus were combined. \textit{Early} switches to BC occurred within the first $30$ actions, while \textit{late} ones happened after that, as defined in our prior work  \shortcite{abdelshiheed2022assessing,abdelshiheed2020metacognition}. The \textit{Ctrl} and \textit{CDL} students did not have problems presented in BC and hence had one less choice. Bonferroni-corrected $(\alpha=.05/3)$ chi-square tests showed that \textit{DRL} $>$ \textit{CDL} $>$ \textit{Ctrl} in their early-switch choices\footnote{`use presented BC' was excluded from pairwise comparisons.}. For example, \textit{DRL} switched early significantly more than \textit{CDL}: $\chi^2 (2,\, N=794) = 54.9, \, \mathit{p}<.0001$. The last five columns in Figure \ref{fig:switch} display the \textit{DRL} students' decisions per training level. Pairwise Bonferroni-adjusted $(\alpha=.05/10)$ chi-square tests revealed that \textit{DRL} early-switch choices significantly increased in the last two levels. For example, \textit{DRL} students switched early in the fourth level significantly more than the third one: $\chi^2 (3,\, N=276) = 25.1, \, \mathit{p}<.0001$.

In essence, while the DRL policy intervened significantly less in the last two training levels (Table \ref{policyDecisions}), Figure \ref{fig:switch} shows that \textit{DRL} students made significantly more early switches in the same two levels. This suggests that \textit{DRL} students became more \textbf{autonomous} as training proceeded. Additionally, \textit{DRL} students switched early more frequently than \textit{Ctrl} and \textit{CDL}.

\subsection{Policy Adaptation for Metacognitive Development}

One of our objectives was to investigate whether the DRL policy adapted to the student's metacognitive knowledge as it changes through logic training. We leveraged association rule mining to observe frequent patterns by the DRL policy. Specifically, for a student, each two consecutive policy decisions were encoded into a transaction represented as $\{a_{t},\, c_{t},\, a_{t+1}\}$, where $a_{t},\, a_{t+1} \in \{Nud,\, Prs,\, No\}$ and refer to the current and next policy decisions, respectively; $c_{t} \in \{Agree,\, Disagree\}$ and indicates whether the student \emph{complied} (agreed/disagreed) with the current policy decision, where agreement and disagreement are defined as:

\vskip -0.1in
\begin{itemize}
    \setlength\itemsep{0.005em} 
    \item \emph{Nudge (Nud)}: early switches to BC represent agreement, while late switches to BC or using FC means disagreement.
    \item \emph{Present in BC (Prs)}: using BC and FC denote agreement and disagreement, respectively.
    \item \textit{No Intervention (No)}: agreement is to use FC, while using BC reflects disagreement.
\end{itemize}

\noindent We focused on rules in the form of $\{a_{t},\, c_{t}\} \Rightarrow a_{t+1}$ to extract meaningful association rules for the next policy decision based on the student's compliance with the current decision. Since $c_{t}$ has two possible values and $a_{t},\, a_{t+1}$ each has three possible values, there are $18$ possible rules. The \emph{support} and \emph{confidence} of each rule were traditionally computed as:

\vskip -0.1in

\begin{align*}
\text{Sup}(\{a_{t}, c_{t}\} \Rightarrow a_{t+1} ) &= \frac{count(\{a_{t}, c_{t}, a_{t+1}\})}{Total} \\
\text{Conf}(\{a_{t}, c_{t}\} \Rightarrow a_{t+1} ) &= \frac{count(\{a_{t}, c_{t}, a_{t+1}\})}{count(\{a_{t},c_{t}\})}
\end{align*}

\noindent where $Total = 598$ [$46$ \textit{DRL} students * $13$ training decisions]. Table \ref{associationRules} lists the top six rules for the DRL policy sorted by their support in descending order. The rules learned from \textit{Decl}$_{DRL}$ and \textit{Proc}$_{DRL}$ students were similar and thus were combined. The remaining rules were excluded due to their significantly low support and confidence values.



\begingroup
\renewcommand{\arraystretch}{1.1}
\begin{table}[ht!]
\begin{center} 
\caption{Top Association Rules for DRL policy} 
\label{associationRules} 
\begin{tabular}{c|c|cc} 
\Xhline{4\arrayrulewidth}

Rank & Rule & Supp ($\%$) & Conf ($\%$)\\
\hline
1 & $\{No,\, Disagree\} \Rightarrow No $ & $23$ & $76$\\

2 & $\{Nud,\, Agree\} \Rightarrow No $ & $12$ & $61$\\

3 & $\{Prs,\, Agree\} \Rightarrow No $ & $10$ & $58$\\

4 &$\{No,\, Agree\} \Rightarrow Nud $ & $10$ & $60$\\

5 &$\{Nud,\, Disagree\} \Rightarrow Prs $ & $7$ & $69$\\

6 & $\{Prs,\, Disagree\} \Rightarrow Nud $ & $5$ & $66$\\

\Xhline{4\arrayrulewidth}
\end{tabular} 
\end{center} 
\end{table}
\endgroup

\subsubsection{Interpreting Association Rules} Table \ref{associationRules} reveals unique perspectives of the DRL policy. In essence, the first \textbf{three} rules suggest that the policy treated those who knew in advance \emph{how} and \emph{when} to use BC as \textit{Conditional} students by \emph{avoiding} interventions in such situations. The last \textbf{two} rules reflect swapping the interventions once a student disagrees with their utility. This finding suggests that students' metacognitive knowledge changes during training and confirms our prior results that metacognitive interventions have different effects on \textit{Declarative} and \textit{Procedural} students \shortcite{abdelshiheed2022mixing}. The \textbf{fourth} rule shows that the DRL policy preferred recommending rather than imposing BC for students who previously used FC.   

\section{Discussions \& Conclusions}

\subsubsection{Bridging the Gap} We showed that our DRL policy caused students with low incoming metacognitive knowledge (declarative and procedural) to outperform their conditional peers, who had the highest knowledge and received no interventions. In other words, DRL bridged the metacognitive knowledge gap between students on a logic tutor, where the interventions were provided, and on a subsequent probability tutor, where students received no interventions. 

\subsubsection{Preparation for Future Learning} Experimental declarative and procedural students received DRL-based interventions on logic and surpassed their no-intervention control peers on logic and probability. This suggests that DRL prepared students for future learning, as they outperformed control students on probability based on logic interventions.

\subsubsection{Autonomy and  Metacognitive Development} The DRL policy adapted to the back-and-forth metacognitive development between declarative, procedural, and conditional students. Specifically, the association mining rules analyses showed that the DRL policy changed its interventions to adapt to the dynamic metacognitive knowledge of students. Hence, students became more autonomous and made effective strategic decisions, even when DRL intervened significantly less.



\subsubsection{Limitations and Future Work} There are at least two caveats in our work. First, splitting students into experimental and control resulted in relatively small sample sizes. Second, the probability tutor supported only one strategy, which restricted our intervention ability to logic. Future work involves implementing forward chaining on the probability tutor, comparing multiple DRL algorithms for our interventions, and comparing the DRL interventions against a stronger control that receives random interventions.

\section{Acknowledgments}


This research was supported by the NSF Grants:
MetaDash: A Teacher Dashboard Informed by Real-Time Multichannel Self-Regulated Learning Data (1660878), Integrated Data-driven Technologies for Individualized Instruction in STEM Learning Environments (1726550), Generalizing Data-Driven Technologies to Improve Individualized STEM Instruction by Intelligent Tutors (2013502) and CAREER: Improving Adaptive Decision Making in Interactive Learning Environments (1651909).


\bibliographystyle{apacite}

\setlength{\bibleftmargin}{.125in}
\setlength{\bibindent}{-\bibleftmargin}

\bibliography{cogsci2023}

\end{document}